\documentclass[11pt,bm,aps,nofootinbib,amsfonts,amssymb,preprintnumbers
]{revtex4}

 \usepackage{here}

\usepackage[dvipdfmx]{graphicx}
\usepackage{amssymb,amsfonts,amsmath,bm,color,multirow,cases,empheq,hyperref}
\usepackage{mathrsfs}

\usepackage{enumerate}
\usepackage{ulem}
\usepackage{afterpage}

\hypersetup{%
 setpagesize=false,
 bookmarksnumbered=true,%
 bookmarksopen=true,%
 colorlinks=true,%
 linkcolor=blue,%
 citecolor=red}

\newcommand{\fr}[1]{\frac{1}{#1}}

\newcommand{\cL}{{\mathcal L}}

\newcommand{\cout}[1]{}

\begin{document}

\preprint{TTI-MATHPHYS-13}
\title{Stable bound orbits around static Einstein-Gauss-Bonnet black holes}
\author{Ryotaku Suzuki}
\author{Shinya Tomizawa}
\email{sryotaku@toyota-ti.ac.jp}
\email{tomizawa@toyota-ti.ac.jp}
\affiliation{Mathematical Physics Laboratory Toyota Technological Institute\\ Hisakata 2-12-1, Nagoya 468-8511, Japan}
\date{\today}

\begin{abstract}

As is well-known, asymptotically flat, static and spherically symmetric black holes do not admit stable bound orbits of massive/massless particles outside the horizon in  higher-dimensional Einstein gravity.  
However, for massive particles, this dramatically changes  in higher curvature theories.
 We clarify the parameter range such that there exist stable bound orbits in $d$-dimensional Einstein-Gauss-Bonnet theories for $6\le d\le 9$. 
In particular, we show the existence of the lower bound of the Gauss-Bonnet coupling constant below which stable bound orbits cease to  exist. 
Moreover,  we also find that for AdS-black holes in the theories, there can exist two stable circular orbits  outside the horizon.

\end{abstract}

\maketitle

\section{Introduction}

A stable bound orbit is an orbit in which a particle keeps moving in a bounded spatial region without reaching infinity or singularities when small perturbations are done. 
Such an orbit often appears in astrophysical phenomena.
For instance, the existence of an innermost stable circular orbit (ISCO) plays an essential role in the formation of the accretion disk around a black hole. 
One can understand the essence of this mechanism by considering the motions of a massive particle moving around a Schwarzschild black hole. 
The massive particle moving along a stable bound orbit is localized near a radial potential well, 
which is made by a balance between the gravitational potential $-Mm/r$ and the centrifugal potential $\ell^2/(m r^2)$, where $M$ is the black hole mass, $m$ and $\ell$ are the mass and angular momentum of a particle, and $r$ is the circumferential radius.
The relativistic correction effect $-M\ell^2/(mr^3)$ becomes dominant over the other effects in the neighborhood of the event horizon and causes proper relativistic phenomena such as perihelion shift and ISCO.
Similarly, such a stable bound orbit appears even around a rotating black hole such as a Kerr background~\cite{Wilkins:1972rs}.

\medskip
On the other hand, 
it is suggested that the existence of stable bound orbits for massless particles may cause nonlinear instability~\cite{Cardoso:2014sna}.
For Schwarzschild spacetime and Kerr spacetime, it is well-known that there exists an unstable circular orbit but not stable one, whereas for the Kerr-Newman spacetime with relatively large electric charge, a stable photon orbit exists on the horizon~\cite{Khoo:2016xqv}. 
Moreover, for the Majumdar-Papapetrou spacetimes with two black holes, such orbits appear even outside the horizon~\cite{Dolan:2016bxj, Nakashi:2019mvs, Nakashi:2019tbz}.
 From the wave perspective, linear waves localize in the vicinity of the trapping null geodesics resulting in a long timescale for the decay~\cite{Keir:2014oka}, which also suggests the existence of nonlinear instabilities of the background spacetime~\cite{Cardoso:2014sna}.

\medskip
 It is also notable that the particle motion drastically changes  in  $d\ (\ge 5)$-dimensions because the gravitational potential and the relativistic correction are replaced with, respectively, $-Mm/r^{d-3}$ and $-M\ell^2/(mr^{d-1})$, while the centrifugal potential remains the four-dimensional one.
For the Schwarzschild-Tangherlini solutions and Myers-Perry solutions in $d=5$, it is shown that for massive particles as well as massless particles, there are no stable circular orbits in equatorial planes~\cite{Tangherlini:1963bw,Frolov:2003en,Page:2006ka,Frolov:2006pe,Cardoso:2008bp}.
In contrast, for the black ring solutions~\cite{Igata:2010ye,Igata:2010cd,Igata:2013be} and the black lens solutions~\cite{Tomizawa:2019egx,Tomizawa:2020mvw} having non-spherical horizon topologies,  it should be surprising that  stable bound orbits for both massive and massless particles exist. 
In addition,
for various types of black holes such as multi-black holes~\cite{Igata:2020vlx} and 
Kaluza-Klein black holes~\cite{Igata:2021wwj,Tomizawa:2021vaa}, the existence of stable bound orbits was also investigated.
Recently, it was shown that there exist stable bound orbits even in microstate geometries, which are the horizonless geometries that can be regarded as a certain kind of approximation of black hole geometries~\cite{Eperon:2016cdd,Eperon:2017bwq,Tomizawa:2022kpt}.

\medskip

Moreover, it is surprising that in higher curvature theories, even asymptotically flat, static spherically symmetric black hole backgrounds admit stable bound orbits for massive particles  in some cases,
unlike the Schwarzschild-Tangherlini black holes in any dimensions of $d\ge 5$.  
It has been clarified that black holes in the $d$-dimensional pure Gauss-Bonnet (GB) theories $(6\le d\le 8)$ and the pure $N$-th Lovelock theories ($2N+2\leq d \leq 4N+4$) allow stable bound orbits~\cite{Dadhich:2013moa,Dadhich:2021vdd}\footnote{Interestingly, a similar dimensionality bound in the massive dust collapse has been shown in the EGB theory~\cite{Maeda:2006pm}.}.
However, the appearance of stable bound orbits in the Einstein-Gauss-Bonnet (EGB) and Einstein-Lovelock theories is not yet fully understood. 
Asymptotically flat, static and spherically symmetric EGB black holes~\cite{Boulware:1985wk} do not have stable bound orbits in $d=5$~\cite{Bhawal:1990nh}, but they do have them in $6 \le d \le 9$ regardless of the presence of a cosmological constant~\cite{Rosa:2008dh}\footnote{In ref.~\cite{Rosa:2008dh}, the authors also claimed to find the stable bound orbits in $d=5$ static EGB solutions, which is contrary to the result in ref.~\cite{Bhawal:1990nh}.
This is because the former has paid no attention to whether the solution describes a black hole spacetime or a horizonless spacetime with naked singularities, which depends on the parameters. 
As a result, their $d=5$ result happened to be in the parameter range of naked singularities.
}
, which apparently seems inconsistent with the nonexistence of stable bound orbits in the $d=9$ pure GB theory~\cite{Dadhich:2013moa,Dadhich:2021vdd}.

In this paper, we aim to clarify the effect of the GB correction and the dimension dependence for the existence of stable bound orbits of static EGB black holes. 
We see that stable bound orbits exist for $6 \le d \le 9$ EGB black holes if the GB coupling constant is sufficiently large compared to the horizon radius. 
In $d=9$, we observe that stable bound orbits have larger radii as the coupling grows, and
 finally, goes to infinity as the GB coupling goes to infinity, consistent with the pure GB result. 
We also find two stable circular orbits when the negative cosmological constant is included.

\medskip

The remaining part of this paper is organized as follows: 
In Sec.~\ref{sec:setup}, we give a brief review of the GB black holes and explain our formalism used in this paper.
In Sec.~\ref{sec:largealpha}, we analytically study the particle motion around EGB black holes with a large enough GB coupling constant. 
In Sec.~\ref{sec:res}, we  discuss the existence of bound orbits  in $6 \le d \le 9$ by numerically investigating the existence of local minima of the effective potential. 
Moreover, in Sec.~\ref{sec:ads}, we also study the effect on the particle motion by a negative cosmological constant. Finally, Sec.~\ref{sec:sum} is devoted to the summary of our results and our future works.

\section{setup}\label{sec:setup}
First, we consider asymptotically flat, static and spherically symmetric black holes in $d$-dimensional EGB theory, whose action is given by
\begin{align}
 S = - \fr{16 \pi G}\int (R + \alpha_{\rm GB} \cL_{\rm GB}) d^d x,
\end{align}
where the GB term $\cL_{\rm GB}$ is written as
\begin{align}
 \cL_{\rm GB} = R^2-4 R_{\mu\nu} R^{\mu\nu} + R_{\mu\nu\rho\sigma}R^{\mu\nu\rho\sigma}.
\end{align}
The metric of the EGB black holes~\cite{Boulware:1985wk} is given by
\begin{align}
&ds^2 = - f(r) dt^2 + \frac{dr^2}{f(r)} + r^2 d\Omega_{d-2}^2,
\end{align}
with the metric on the $(d-2)$-dimensional sphere,
\begin{align}
& \qquad d\Omega_{d-2}^2 = d\theta^2 + \sin^2 \theta d\phi^2 + \cos^2\theta d\Omega_{d-4}^2,
\end{align}
the metric function
\begin{align}
f(r) = 1 + \frac{r^2}{2\alpha}\left(1-\sqrt{1+\frac{4\alpha(\alpha+r_0^2)r_0^{d-5}}{r^{d-1}}}\right),\label{eq:EGB-fr}
\end{align}
and the scaled GB coupling constant
\begin{align}
\alpha:=(d-3)(d-4)\alpha_{\rm GB}.
\end{align}

In this paper, we consider only the GR (General Relativity) branch which has an asymptotically flat region. We set the parametrization so that the horizon appears at $r=r_0$ for any values of $\alpha$. The black hole mass is not fixed for $\alpha$ because
\begin{align}
 M_{\rm BH} = \frac{(d-2)\Omega_{d-2}(\alpha+r_0^2)r_0^{d-1}}{16\pi G}.
\end{align}
Thanks to the spherical symmetry, we can set $\theta=\pi/2$ without loss of generality. 
The geodesic motion of massive/massless particles on an equatorial plane $\theta=\pi/2$  is described by the Hamiltonian
\begin{align}
H =  g^{rr} p_r^2 + E^2 g^{tt} +L_\phi^2 g^{\phi\phi} + m^2,
\end{align}
where $p_t=-E,\, p_\phi=L$ are the constants of motion which correspond to the energy and angular momentum of particles, respectively.
Following the Hamiltonian constraint $H=0$,  the particles can move only in the region where the effective potential $V_{\rm eff}$ satisfies 
\begin{align} 
 V_{\rm eff} + m^2/E^2 = - E^{-2} g^{rr} p_r^2 \leq 0
\end{align}
where
\begin{align}
 V_{\rm eff}  := g^{tt}+(L_\phi^2/E^2) g^{\phi\phi} = -\fr{f(r)} +  \frac{\ell^2}{r^2}.
\end{align}
From the symmetry $\phi \to -\phi$ of the solution, we can assume $\ell:=L_\phi/E \geq0$ without loss of generality. 
Stable bound orbits for massive particles exist if the effective potential $V_{\rm eff}$ has a negative local minimum, whereas, those for massless particles exist if it has a finite negative region bounded by $V_{\rm eff}=0$.
In particular, the stable circular orbits (SCOs) at $r=r_{\rm SCO}$ are  given by the local minima of $V_{\rm eff}$,
\begin{align}
 V_{\rm eff}'(r_{\rm SCO}) = \frac{f'(r_{\rm SCO})}{f(r_{\rm SCO})^2}-\frac{2\ell^2}{r_{\rm SCO}{}^3} =0,\quad V_{\rm eff}''(r_{\rm SCO})>0, 
\end{align}
where SCOs for massless particles should  satisfy the condition of $V_{\rm eff}(r_{\rm SCO})=0$.
By further differentiating with $\ell$, one can show that $r_{\rm SCO}$  is a monotonic function of $\ell$,
\begin{align}
\left. \frac{d r_{\rm SCO}}{d\ell}\right|_{\alpha}  =  \frac{4\ell}{r_{\rm SCO}^3V_{\rm eff}'' (r_{\rm SCO})}>0.\label{eq:ell-r_sco_mono}
\end{align}
Thus, the minimum and maximum values of $\ell$ for bound orbits correspond to the ISCO and outermost stable circular orbit (OSCO), respectively.

\section{large $\alpha$ limit}\label{sec:largealpha}
First, we study the stable  bound orbits at the large $\alpha$ limit. More precisely, we assume that $\alpha$ is large in units of the horizon radius
\begin{align}
 \alpha /r_0^2 \gg 1,
\end{align}
which can also be regarded as the approximation of a small black hole  when $\alpha$ is fixed. 
In particular, the effective potential at the pure GB limit $\alpha \to\infty$ behaves like
that of the $(d+1)/2$-dimensional Schwarzschild black hole spacetime,
\begin{align}
V_{\rm eff} \to -\left(1 - \left(\frac{r_0}{r}\right)^{\frac{d-5}{2}}\right)^{-1}+ \frac{\ell^2}{r^2}.
\end{align}
In this limit, as explained in~\cite{Dadhich:2013moa,Dadhich:2021vdd} (see Table~\ref{table:2}), the power $(d-5)/2$ of the gravitational potential is smaller than the power $2$ of the centrifugal potential  only for $6 \le d \le 8$, which gives the potential a local minimum for these dimensions. 
As was shown  in the $N$-th pure Lovelock theory for $d=2N+2$~\cite{Dadhich:2013moa}, which corresponds to the pure GB case for $d=6$, the minimum values of $\ell$ and $r_{\rm ISCO}$ are obtained for $6 \le d \le 8$ by solving the conditions $V_{\rm eff}'=0$ and $V_{\rm eff}''=0$ as
\begin{align}
& \ell_{\rm min}^{\alpha=\infty} = \frac{(9-d)^\frac{d-9}{2(d-5)}(d-1)^\frac{d-1}{2(d-5)}}{4\sqrt{d-5}} r_0,\\
 & r_{\rm ISCO}^{\alpha=\infty} = \left(\frac{d-1}{9-d}\right)^{\frac{2}{d-5}}r_0.
\end{align}
Furthermore, these values can be smoothly continued to the finite $\alpha$ by the $1/\alpha$ expansion
\begin{align}
 \ell_{\rm min} = \ell_{\rm min}^{\alpha=\infty} (1+c_1 (r_0^2/\alpha)+\dots),
 \quad r_{\rm ISCO} = r_{\rm ISCO}^{\alpha=\infty} (1+c_2 (r_0^2/\alpha)+ \dots),
\end{align}
where the first coefficients for each dimension are given by
\begin{align}
& c_1^{d=6}=\frac{4097}{972}, \quad  c_1^{d=7}=\frac{29}{7},  \quad c_1^{d=8}=\frac{1}{3}+\frac{49}{4}7^{1/3}\\
&  c_2^{d=6}=-\frac{14896}{729},\quad c_2^{d=7}=-40, \quad c_2^{d=8}=\frac{1}{3}-\frac{343}{3} 7^{1/3}.
\end{align}
Since for $d\geq 9$, the potential behaves like that of the Schwarzschild in $d\geq 5$ even at the pure GB limit,
one may not be able to expect the existence of stable bound orbits. 
However, as will be seen later, the potential admits the existence of stable  bound orbits  even in $d=9$ for large and finite $\alpha$, though it does not do so for the pure GB limit $\alpha\to\infty$.

\begin{table}[H]

  \centering
  \begin{tabular}{|c|c|c|}
    \hline
$\  d\ $& Schwarzschild BH  & Gauss-Bonnet BH\\\hline \hline
5  & 2   & 0  \\
6  & 3   & 1/2  \\
7  & 4   & 1  \\
8  & 5   & 3/2  \\
9  &  6  & 2  \\
 \hline
  \end{tabular}
  \caption{Powers of gravitational potentials for $d$-dimensional Schwarzschild black holes and $d$-dimensional GB black holes.    \label{table:2}}
\end{table}

Next, we consider the case in which $\alpha$ is large but still finite, where
the effective potential has an inflection point at which the behavior of the gravitational potential switches from one in lower dimensions to  one in  higher dimensions, or from the potential of pure GB to that of GR.
This transition leads to the existence of a maximum for $r_{\rm SCO}$, namely, the existence of OSCO for $d=9$ as well as $6\le d\le 8$. 
It can be expected qualitatively that the transition happens in the range where both terms in the square root in eq.~(\ref{eq:EGB-fr}) become comparable, 
\begin{align}
 \frac{\alpha^2 r_0^{d-5}}{r^{d-1}} \sim 1,
\end{align}
which gives a transition scale
\begin{align}
r \sim r_{\rm tr} = \left(\frac{\alpha}{r_0^2}\right)^\frac{2}{d-1} r_0.\label{eq:scale-rc}
\end{align}
For $\alpha/r_0^2 \gg 1$, the transition occurs at a distance far enough from the horizon, $r_{\rm tr} \gg r_0$.
For $r \gg r_{\rm tr}$, the effective potential behaves as that of the $d$-dimensional Schwarzschild black hole spacetime, 
\begin{align}
V_{\rm eff} \simeq -1-\frac{(\alpha/r_0^2+1)r_0^{d-3}}{r^{d-3}}+\frac{\ell^2}{r^2},
\end{align}
which does not admit a local minimum for $d\geq 5$.
Thus, for $d=6,7,8$, the potential behaves as
\begin{align}
 V_{\rm eff} \sim \left\{
 \begin{array}{cc}
- 1- (r/r_0)^{-\frac{d-5}{2}} & (r_0 \ll r \ll r_{\rm tr})\\
 -1+\ell^2 r^{-2} & (r\gg r_{\rm tr})
 \end{array}\right.,
\end{align}
which guarantees the local maximum across $r\sim r_{\rm tr}$. 
Moreover, it is notable that this transition still occurs even for $d=9$, though we need a slight modification, since the potential behaves as 
\begin{align}
 V_{\rm eff} \sim \left\{
 \begin{array}{cc}
- 1+ (\ell^2-r_0^2) r^{-2} & (r_0 \ll r \ll r_{\rm tr})\\
 -1+\ell^2 r^{-2} & (r\gg r_{\rm tr})
 \end{array}\right.,
\end{align}
which again guarantees the local maximum if $\ell \lesssim r_0$.
 Since as previously mentioned, a larger angular momentum yields a larger stable  circular orbit radius~(\ref{eq:ell-r_sco_mono}),
the orbit has maximum $\ell$ when $r_{\rm SCO}$ reaches this maximum.
To roughly estimate the maximum around $r\sim r_{\rm tr}$, we expand the effective potential in the power series of $r_0^2/\alpha$,
\begin{align}
V_{\rm eff} \simeq -1 + \left(\frac{\alpha}{r_0^2}\right)^{-\frac{d-5}{d-1}} U_{\ell/\ell_{\rm tr}}(r/r_{\rm tr}) 
\end{align}
where the leading order contribution is given in the scale-invariant form
\begin{align}
  U_{\hat\ell}(\hat{r}) = \frac{\hat{r}{}^2}{2}\left(1-\sqrt{1+\frac{4}{\hat{r}{}^{d-1}}}\right)+\frac{\hat{\ell}{}^2}{\hat{r}{}^2},
\end{align}
and the typical transition scale for the angular momentum is also given by
\begin{align}
 \ell_{\rm tr} :=  \left(\frac{\alpha}{r_0^2}\right)^\frac{9-d}{2(d-1)}r_0.
\end{align}
In some parameter range, $U_{\hat{\ell}}(\hat{r})$ admits a local minimum and a local  maximum. The maximum value of such $\hat{\ell}$ is given by solving $U_{\hat{\ell}}'(\hat{r})=U_{\hat{\ell}}''(\hat{r})=0$, which also determines $\hat{r}_{\rm OSCO}$. 
For $6\le d \le 9$, the values of $\hat{\ell}_{\rm max}$ and $\hat{r}_{\rm OSCO}$ are listed in Table~\ref{table:osco}.
\begin{table}[H]
 \centering
  \begin{tabular}{|c||c|c|c|c|}
   \hline
 $d$  & $6$ & $7$ & $8$ & $9$  \\ \hline
$\hat{\ell}_{\rm max}$ 
&  0.869   &  0.903&  0.980&  1.086\\ \hline
$\hat{r}_{\rm OSCO}$  &  1.569  &  1.238  &  1.072  &  0.942\\
\hline
      \end{tabular}
 \caption{Values of $\hat{\ell}_{\rm max}$ and $\hat{r}_{\rm OSCO}$ for $6\le d \le 9$. \label{table:osco}}
\end{table}
\noindent
Therefore, for large $\alpha$, $\ell_{\rm max}$ and $r_{\rm max}$ behave, respectively, as
\begin{align}
 \ell_{\rm max} \propto \ell_{\rm tr} \propto \alpha^{\frac{9-d}{2(d-1)}} ,\quad r_{\rm OSCO} \propto r_{\rm tr} \propto \alpha^\frac{2}{d-1}.
\end{align}

\section{Bound orbits for $6\le d \le 9$}\label{sec:res}
\begin{figure}[H]
\begin{center}
\includegraphics[width=4.cm]{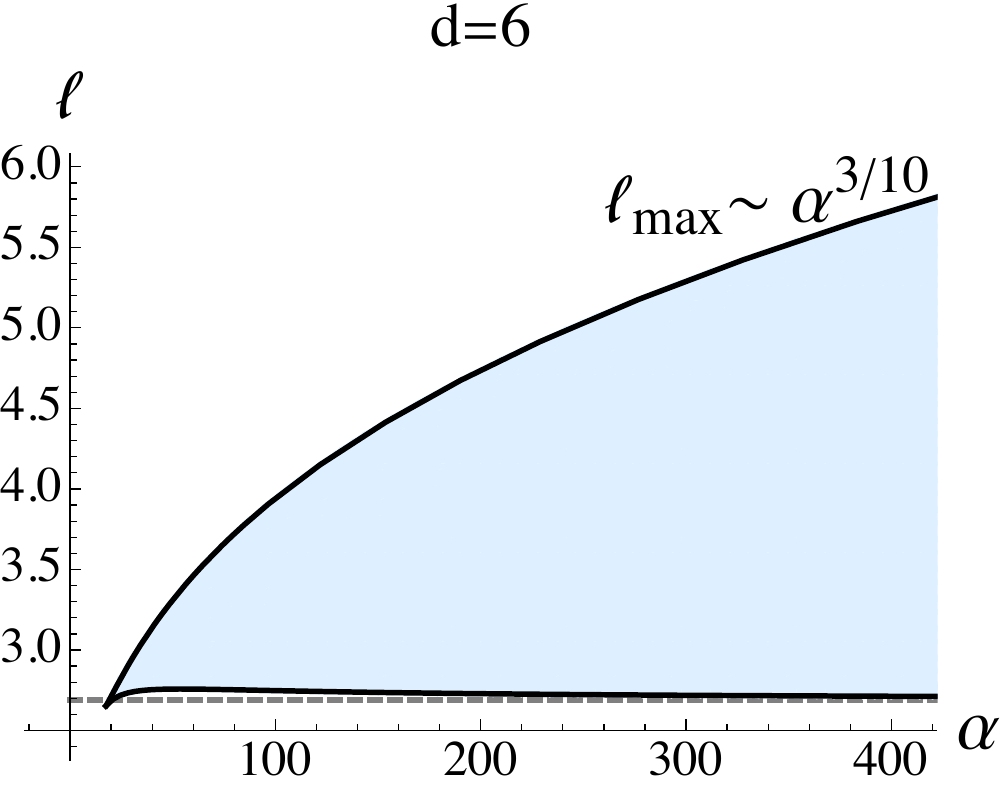}
\includegraphics[width=4.cm]{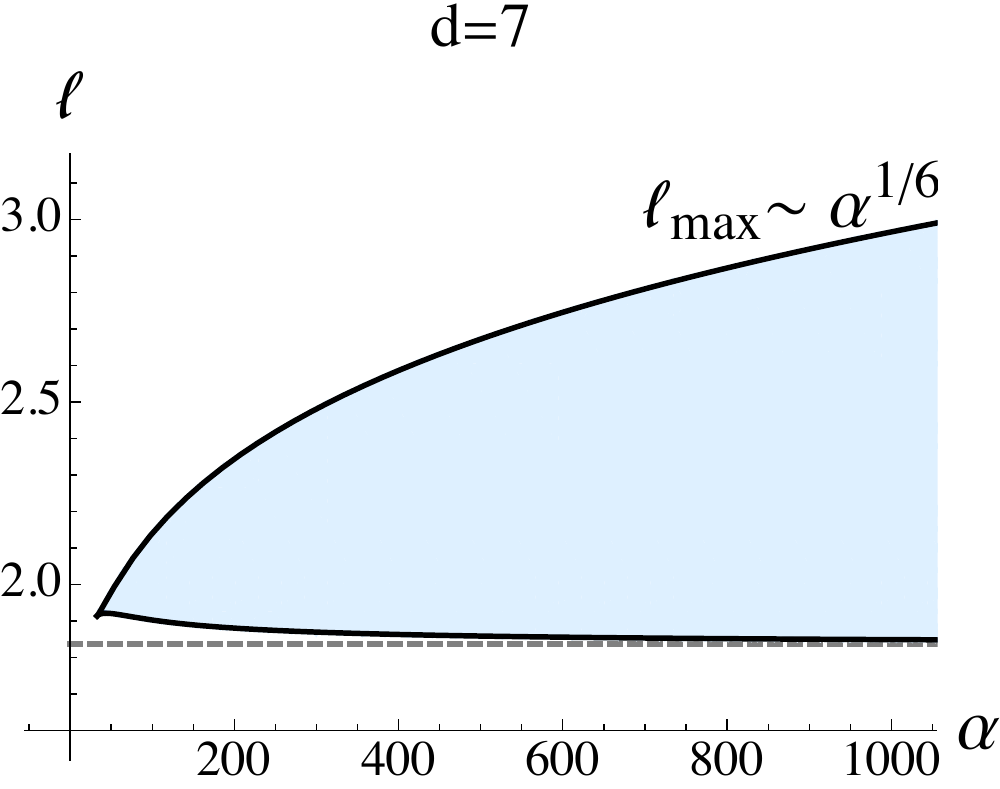}
\includegraphics[width=4.cm]{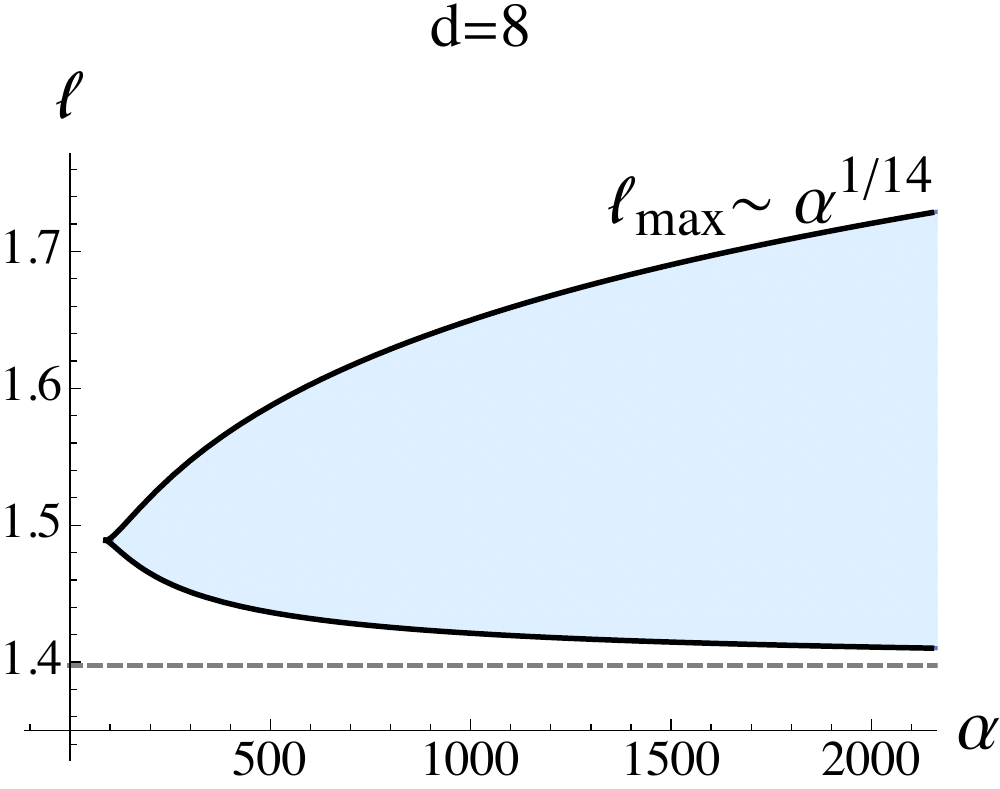}
\includegraphics[width=4.cm]{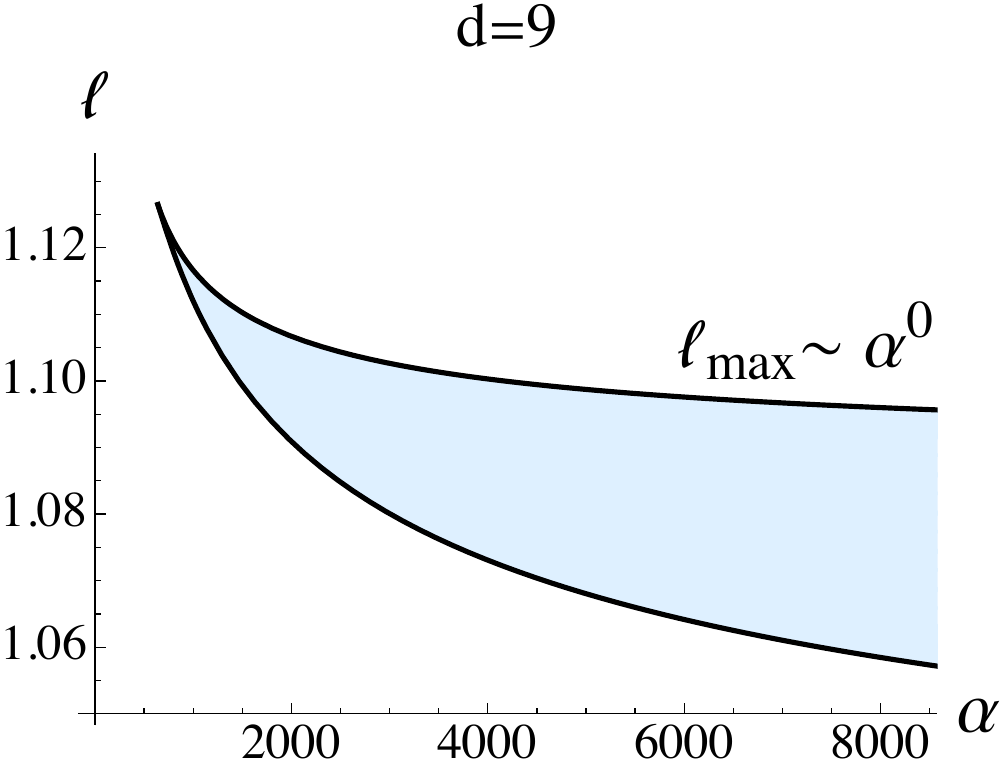}
\end{center}
\caption{Stable bound orbits, shown in the colored region. Note that $\ell$ and $\alpha$ are normalized in units of $r_0$. The dashed lines represent $\ell=\ell_{\rm min}^{\alpha = \infty}$.\label{fig:SCOplot}}
\end{figure}

\begin{figure}[H]
\begin{center}
\includegraphics[width=4cm]{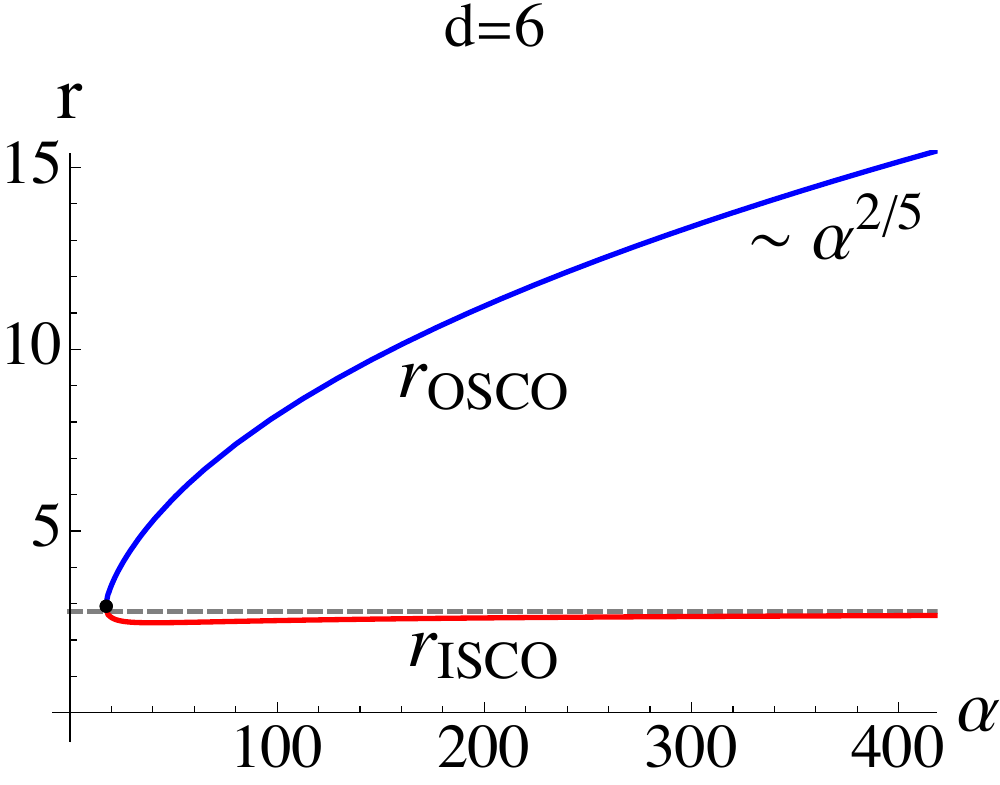}
\includegraphics[width=4cm]{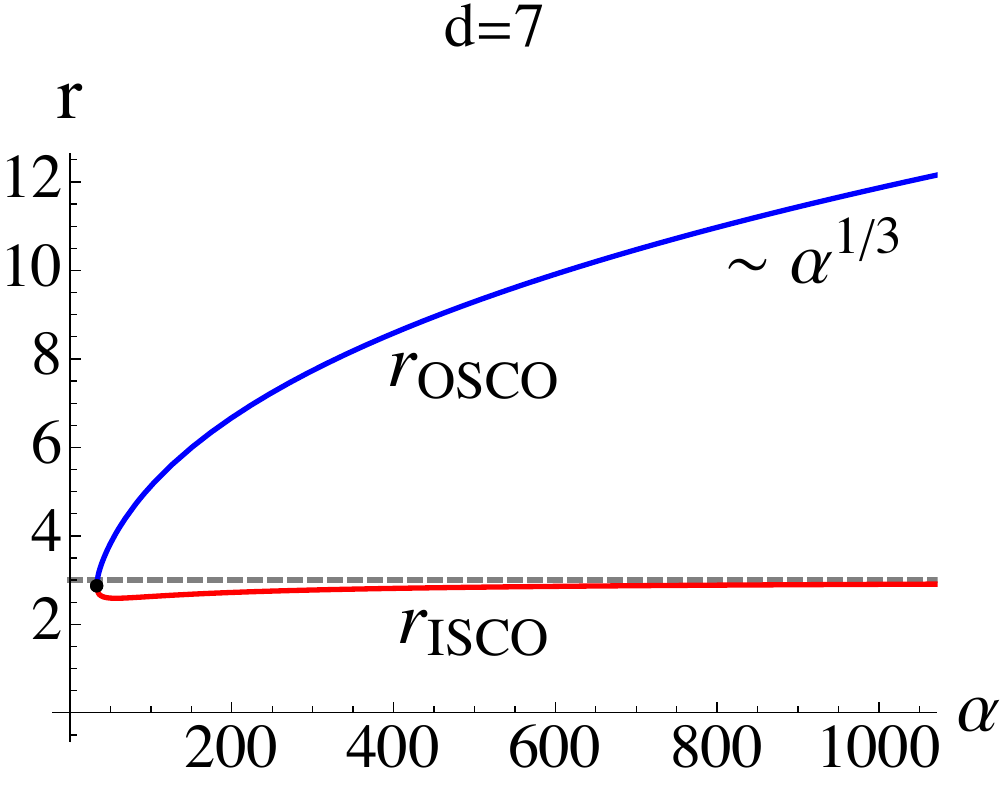}
\includegraphics[width=4cm]{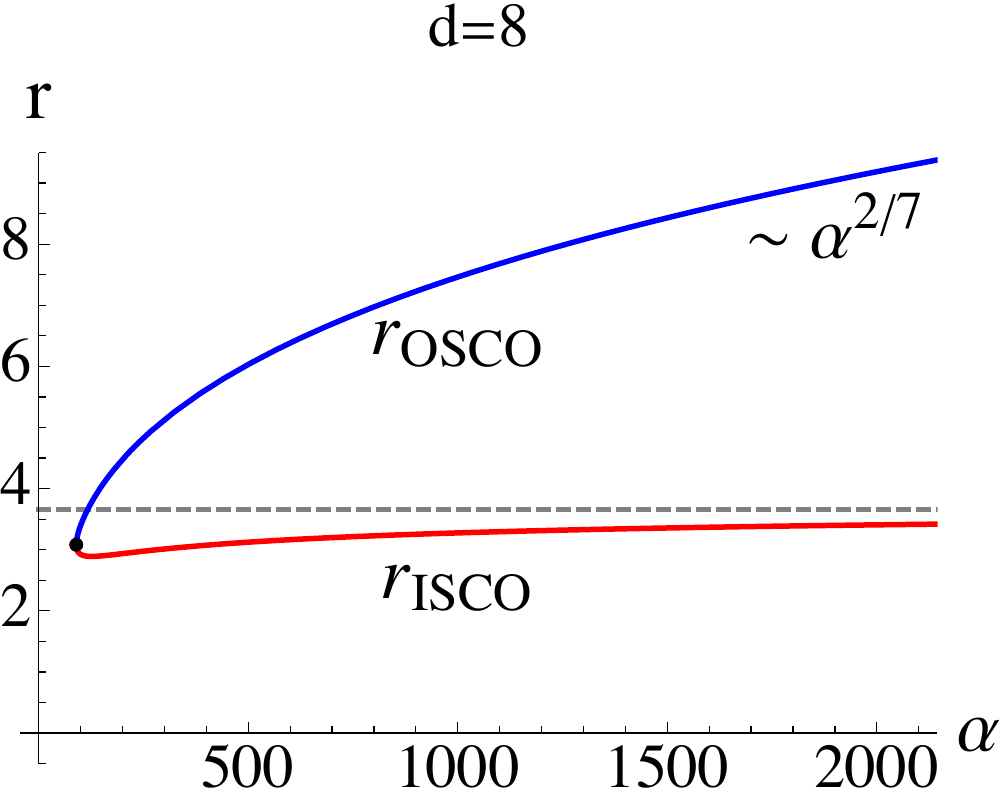}
\includegraphics[width=4cm]{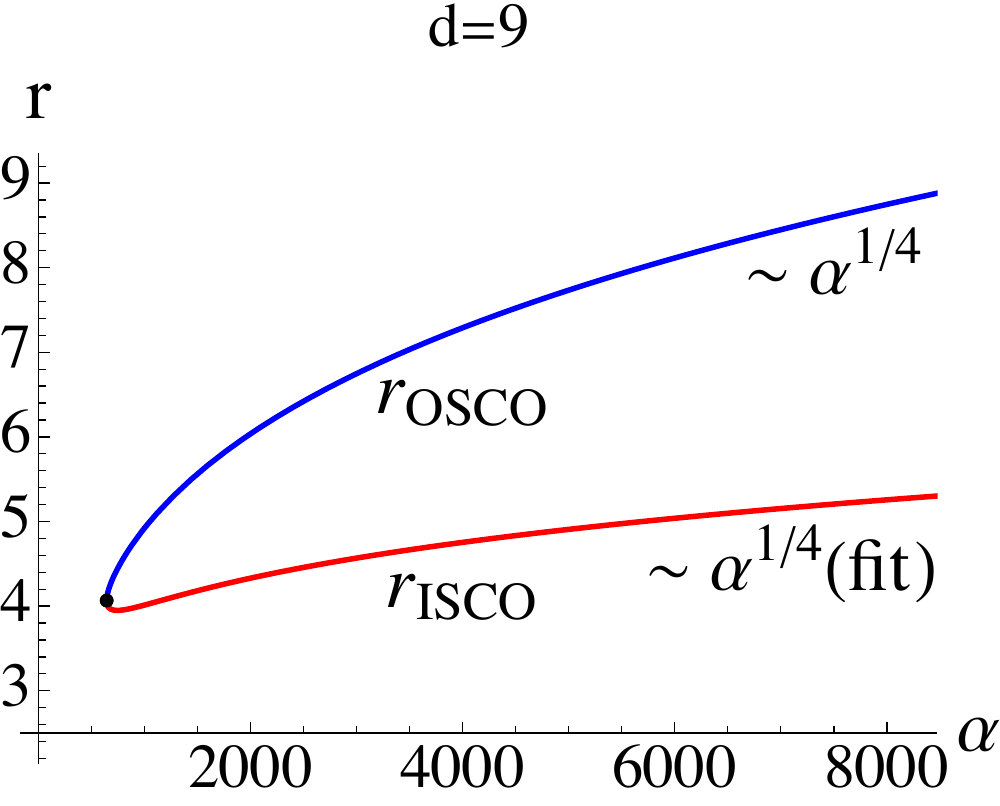}
\end{center}
\caption{Position of ISCO and OSCO for each $\alpha$. Note that $r$ and $\alpha$ are normalized in units of $r_0$. The dashed lines represent $r=r_{\rm ISCO}^{\alpha=\infty}$.\label{fig:rSCOplot}}
\end{figure}

Figures~\ref{fig:SCOplot} and \ref{fig:rSCOplot} show the parameter range of $\ell$ for the existence of  stable bound orbits and the radii of ISCO/OSCO for given $\alpha$, respectively. The curves for $\ell=\ell_{\rm min/max}(\alpha)$ and $r=r_{\rm ISCO/OSCO}(\alpha)$ are numerically determined by solving $V_{\rm eff}'(r)=V_{\rm eff}''(r)=0$.
 In any dimension, we find no stable bound orbits when $\alpha$ is below a critical value $\alpha_c$. The large $\alpha$ behavior of $\ell_{\rm min/max}$ and $r_{\rm ISCO/OSCO}$ is consistent with the analytic result from the large $\alpha$ limit. 
Although we could not find the analytical limit for $\ell_{\rm min}$ and $r_{\rm ISCO}$ in $d=9$, the numerical fit implies $r_{\rm ISCO}\propto \alpha^{1/4}$ at large $\alpha$, which indicates the stable bound orbits disappear at $\alpha \to \infty$ in $d=9$.

\begin{figure}[H]
\includegraphics[width=4cm]{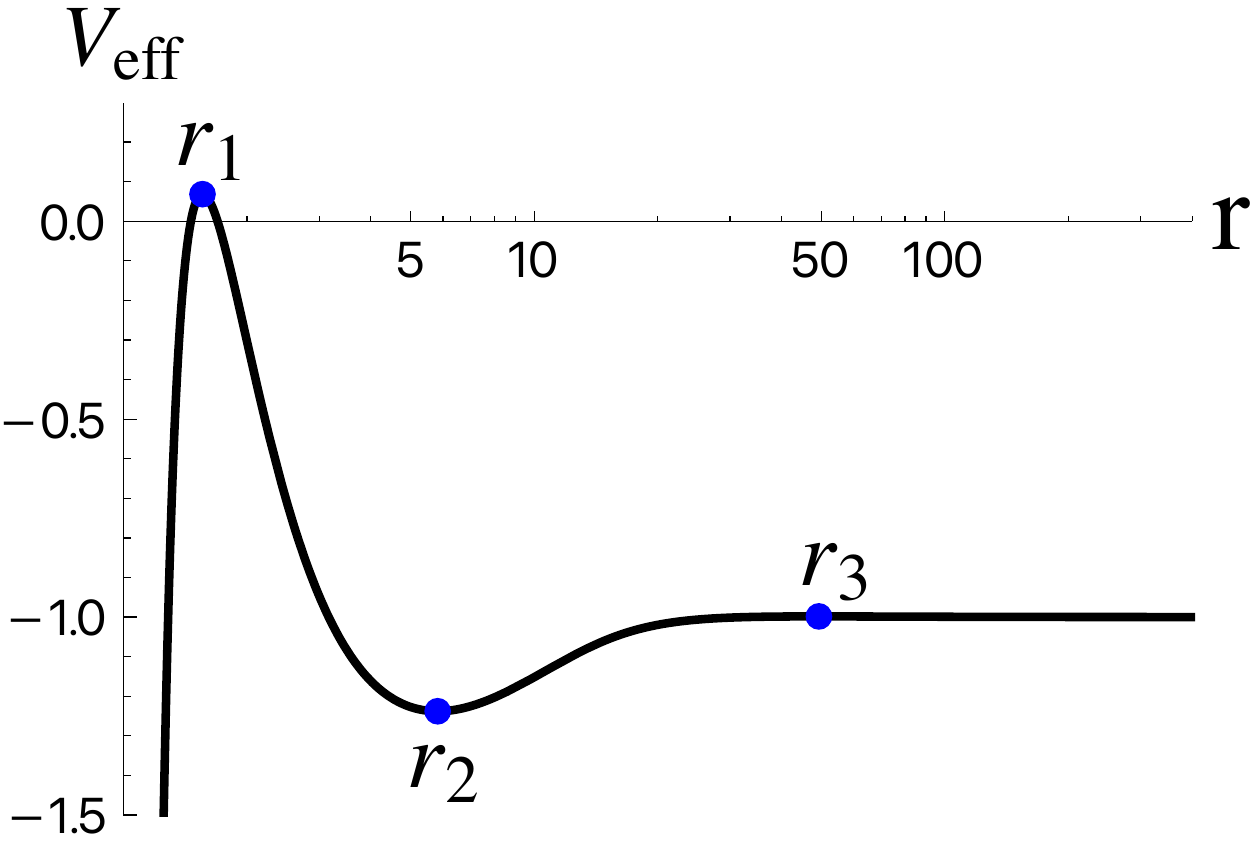}
\includegraphics[width=4cm]{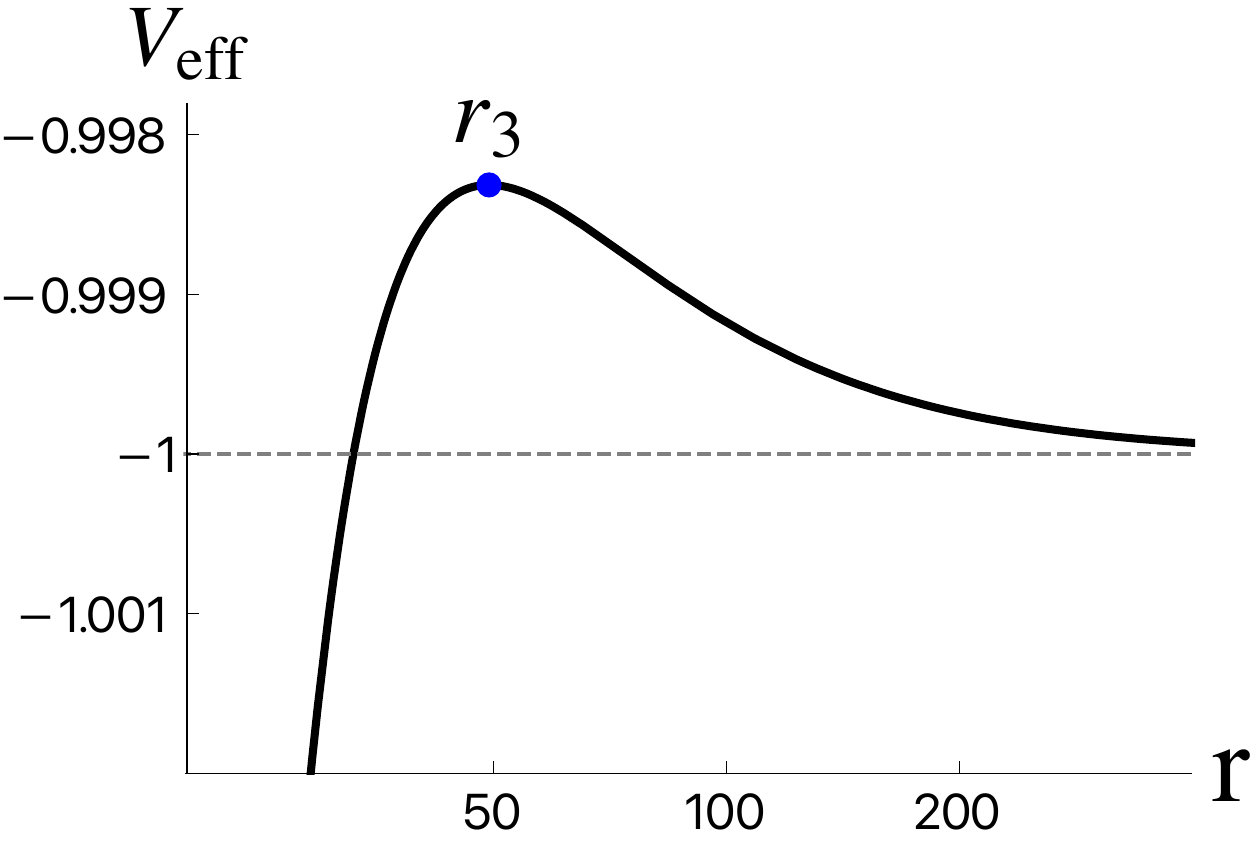}
\includegraphics[width=4cm]{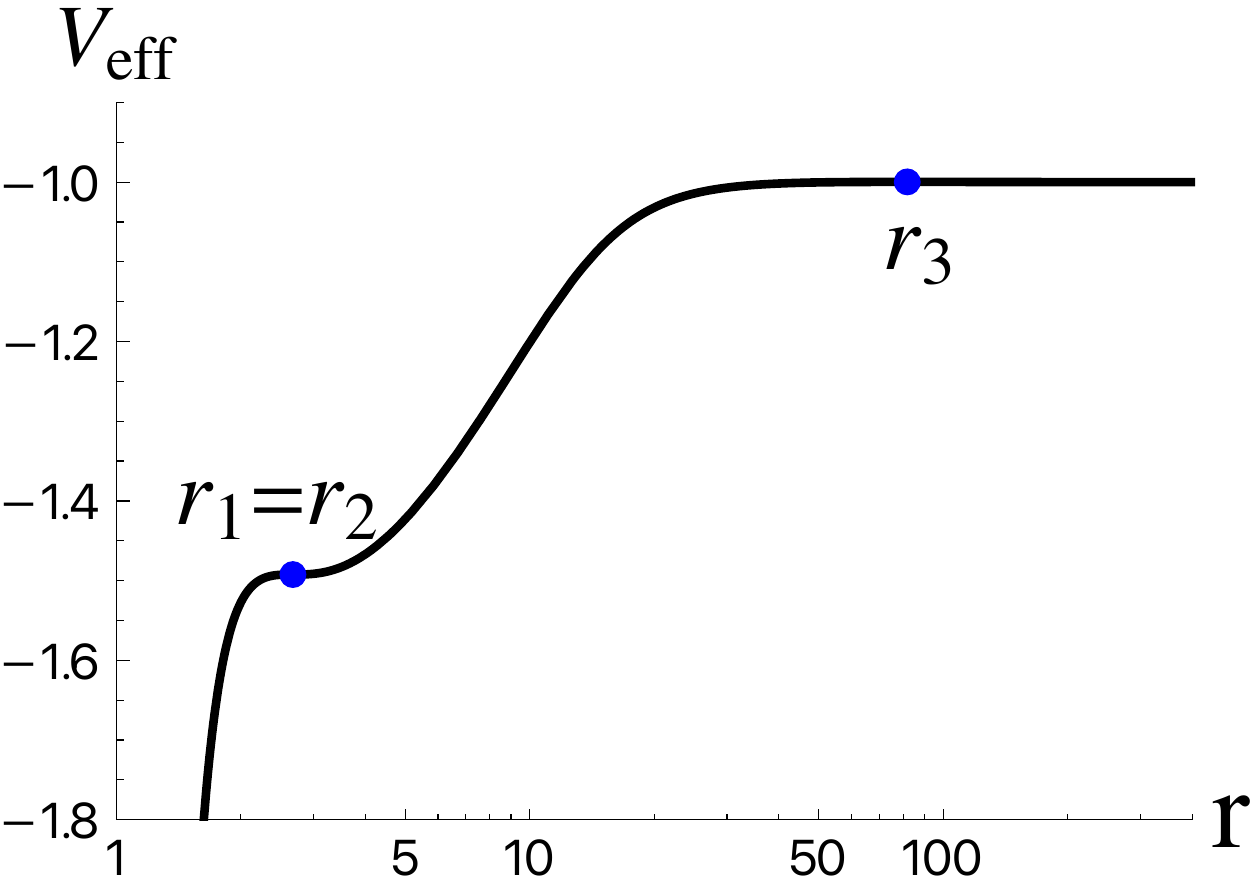}
\includegraphics[width=4cm]{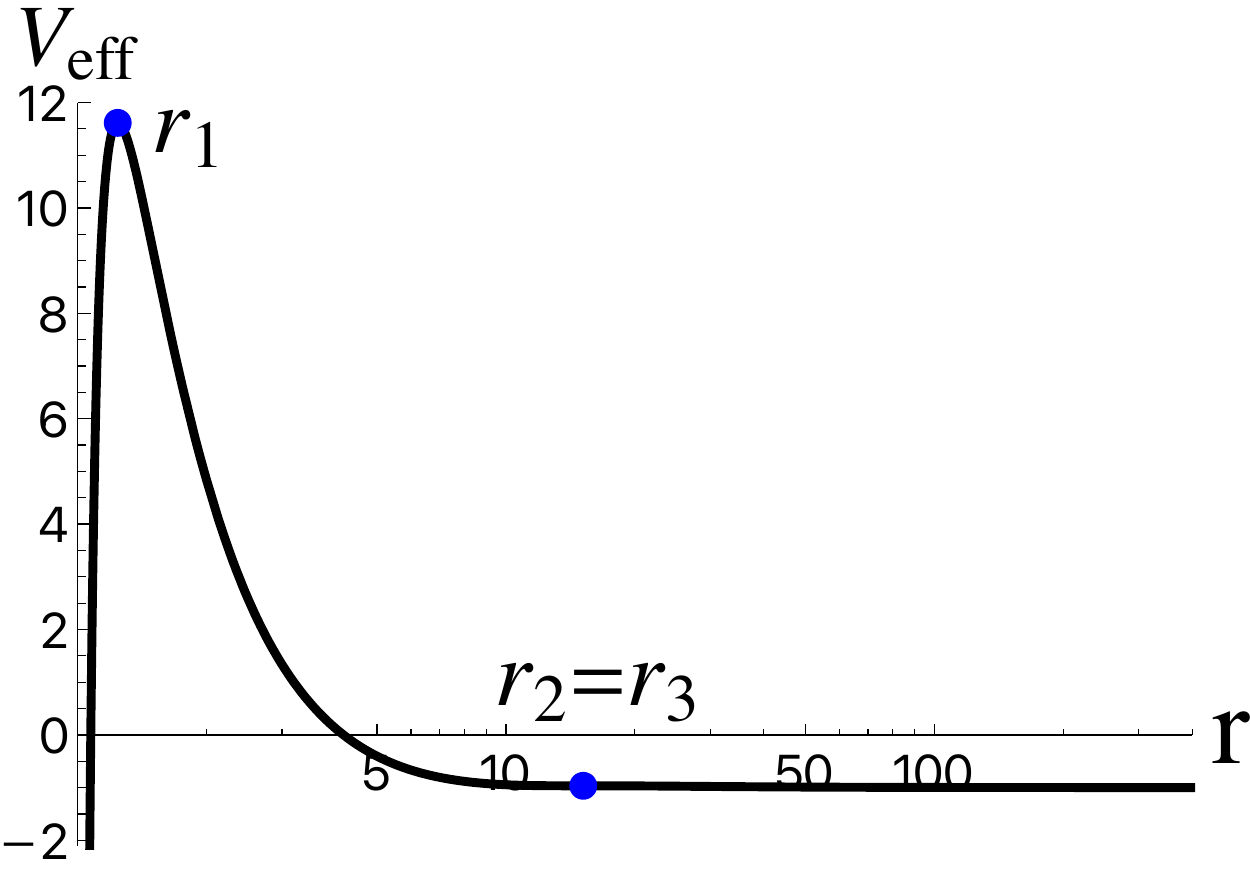}
\caption{Effective potential for $d=6, \alpha=400\, r_0^2$ with $\ell=3.5\, r_0$ (first two panels), $\ell=\ell_{\rm min}\approx 2.71\, r_0$ (third), and $\ell=\ell_{\rm max}\approx 5.73\,r_0$ (fourth). The position is normalized by $r_0$ and plotted in the log scale so that all extrema are visible. The second panel shows the detailed profile around the maximum at $r=r_3$ in the first panel. \label{fig:Veff-1}}
\end{figure}

\begin{figure}[H]
\begin{center}
\includegraphics[width=5cm]{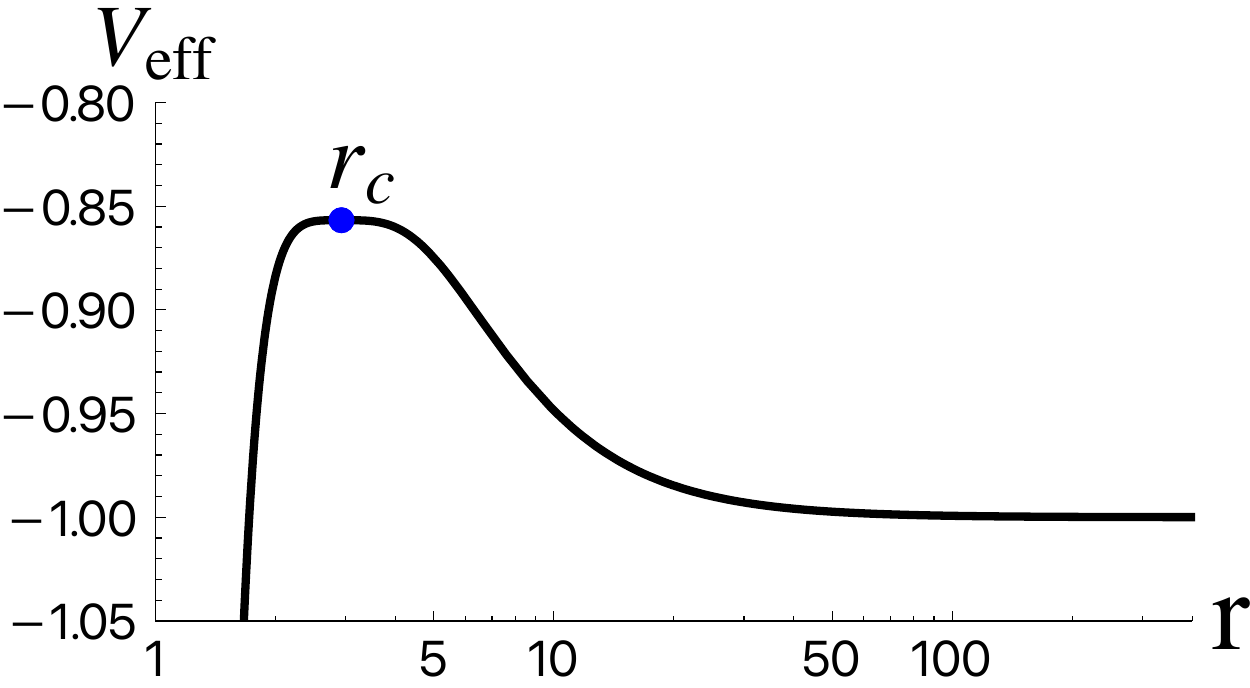}
\caption{Effective potential in $d=6$ with the critical parameteres $\alpha=\alpha_c=17.48 \,r_0^2$ and $\ell=\ell_c=2.652 \,r_0$.\label{fig:Veff_cusp}}
\end{center}
\end{figure}

Figure~\ref{fig:Veff-1} shows typical shapes of the effective potential.
At the limiting value of $\ell$, the potential minimum at $r=r_2$ disappears by merging with either of the local maxima at both sides.
As seen in Fig.~\ref{fig:Veff_cusp},  all three extrema become degenerate at $r_1=r_2=r_3=r_c$ 
 at the critical values $\alpha= \alpha_c$ and $\ell=\ell_{\rm max}=\ell_{\rm min}=\ell_c$ in Table~\ref{table:alpha_c},
 which are determined by
\begin{align}
  V_{\rm eff}'(r_c)\bigr|_{\alpha_c,\ell_c}=V_{\rm eff}''(r_c)\bigr|_{\alpha_c,\ell_c}=V_{\rm eff}^{(3)}(r_c)\bigr|_{\alpha_c,\ell_c}=0.\label{eq:cusp-cond}
\end{align}
This in turn determines the maximum size of the black hole for stable  bound orbits when $\alpha$ is fixed.

\begin{table}[H]
\centering
\begin{tabular}{|c||c|c|c|c|}\hline
$d$ &6 &7&8&9\\\hline
 $\alpha_c/r_0^2$ & 17.48&32.97&89.49&640.6\\\hline
  $V_{\rm eff}(r_c)_{\ell_c,\alpha_c}$ &-0.857&-0.889&-0.930&-0.975
\\\hline
\end{tabular}
\caption{The $\alpha_c$ and corresponding values of $V_{\rm eff}$ at $r=r_c$ for $6\le d \le 9$. 
The latter gives the upper bound of the outer maximum $V_{\rm eff}(r_3)$ in each dimension.\label{table:alpha_c}}
\end{table}

\begin{figure}[H]
\begin{center}
\includegraphics[width=7cm]{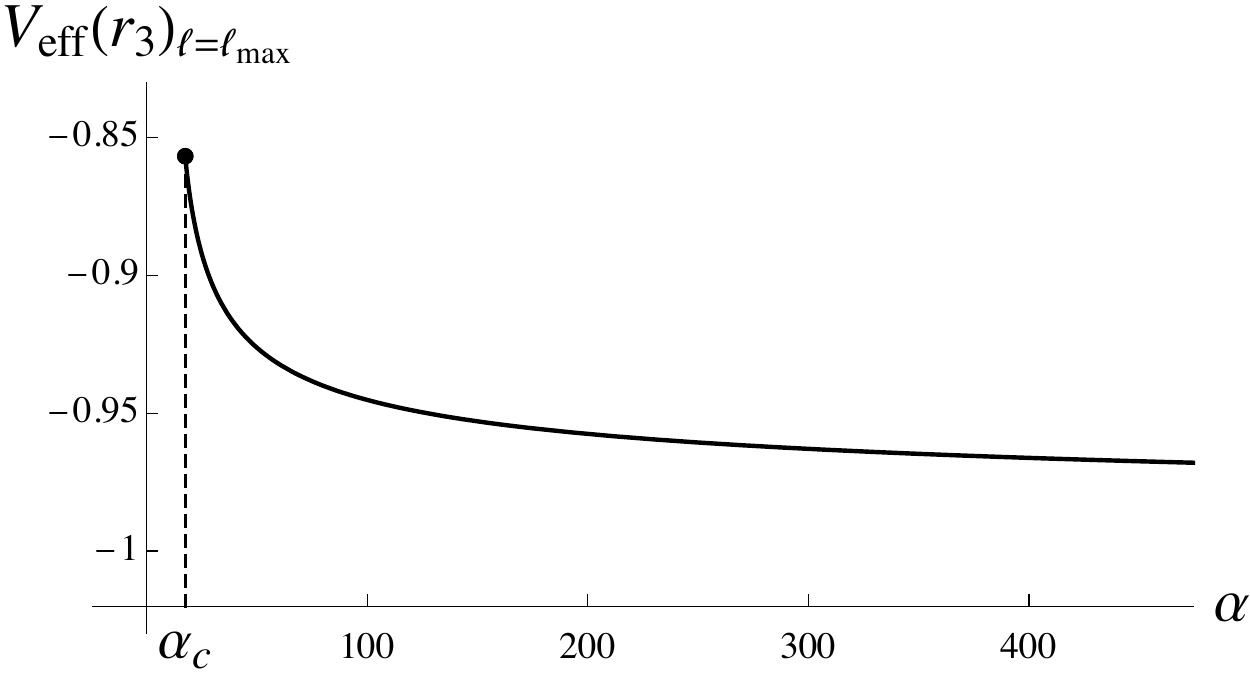}
\caption{The $\alpha$-dependence of $V_{\rm eff}(r_3)$ with $\ell = \ell_{\rm max}$ in $d=6$. Note that $d=7,8,9$ show the same behavior. \label{fig:V3max}}
\end{center}
\end{figure}

Finally, let us evaluate the possible height of the outer maximum $V_{\rm eff}(r_3)$ which determines the maximum energy of the particles trapped in stable bound orbits.
Furthermore, the positivity of $V_{\rm eff}(r_3)$ is the necessary condition for massless particles to have stable bound orbits.
It is easy to show that $V_{\rm eff}(r_3)$ is a monotonic function of $\ell$ for fixed $\alpha$
\begin{align}
 \left.\frac{dV_{\rm eff}(r_3)}{d\ell}\right|_{\alpha} = \frac{2\ell}{r_3^2}> 0,
\end{align}
which leads to $V_{\rm eff}(r_3) \leq V_{\rm eff}(r_3)_{\ell=\ell_{\rm max}}$ for given $\alpha$. In addition, our numerical calculation shows that $V_{\rm eff}(r_c)_{\alpha_c,\ell_c}$ gives the maximum value of $V_{\rm eff}(r_3)_{\ell=\ell_{\rm max}}$ (Fig.~\ref{fig:V3max}), and hence the upper bound for all $V_{\rm eff}(r_3)$.
The values of $V_{\rm max}(r_c)_{\alpha_c,\ell_c}$ for $ 6 \le d \le 9 $ are shown in the bottom row of Table~\ref{table:alpha_c}.
Since the outer maximum is always negative, there are no  stable bound orbits for massless particles.

\section{Two stable circular orbits with $\Lambda<0$}\label{sec:ads}
It is known that  the existence of a negative cosmological constant allows stable  bound orbits around static black holes in Einstein-Lovelock theories~\cite{Konoplya:2020ptx}. 
Therefore, one can expect that the effective potential admits two local minima for $6 \le d \le 9$ , if the increase of the potential due to the AdS barrier occurs sufficiently far away from the transition at $r\sim r_{\rm tr}$ which we discussed in the previous section.

For the EGB-AdS black hole, in terms of the AdS scale $L$ introduced by
\begin{align}
\Lambda = -\frac{(d-1)(d-2)}{2L^2},
\end{align}
eq.~(\ref{eq:EGB-fr}) is merely replaced with
\begin{align}
 f(r) = 1+\frac{r^2}{2\alpha} \left(1-\sqrt{1-\frac{4 \alpha }{L^2}+\frac{4 r_0^{d-3}\alpha  \left(1+\frac{\alpha}{r_0^2} +\frac{r_0^2}{L^{2}}\right)}{ r^{d-1}}}\right),
\end{align}
where the horizon radius is fixed at $r=r_0$.
\begin{figure}[t]
\begin{center}
\includegraphics[width=6cm]{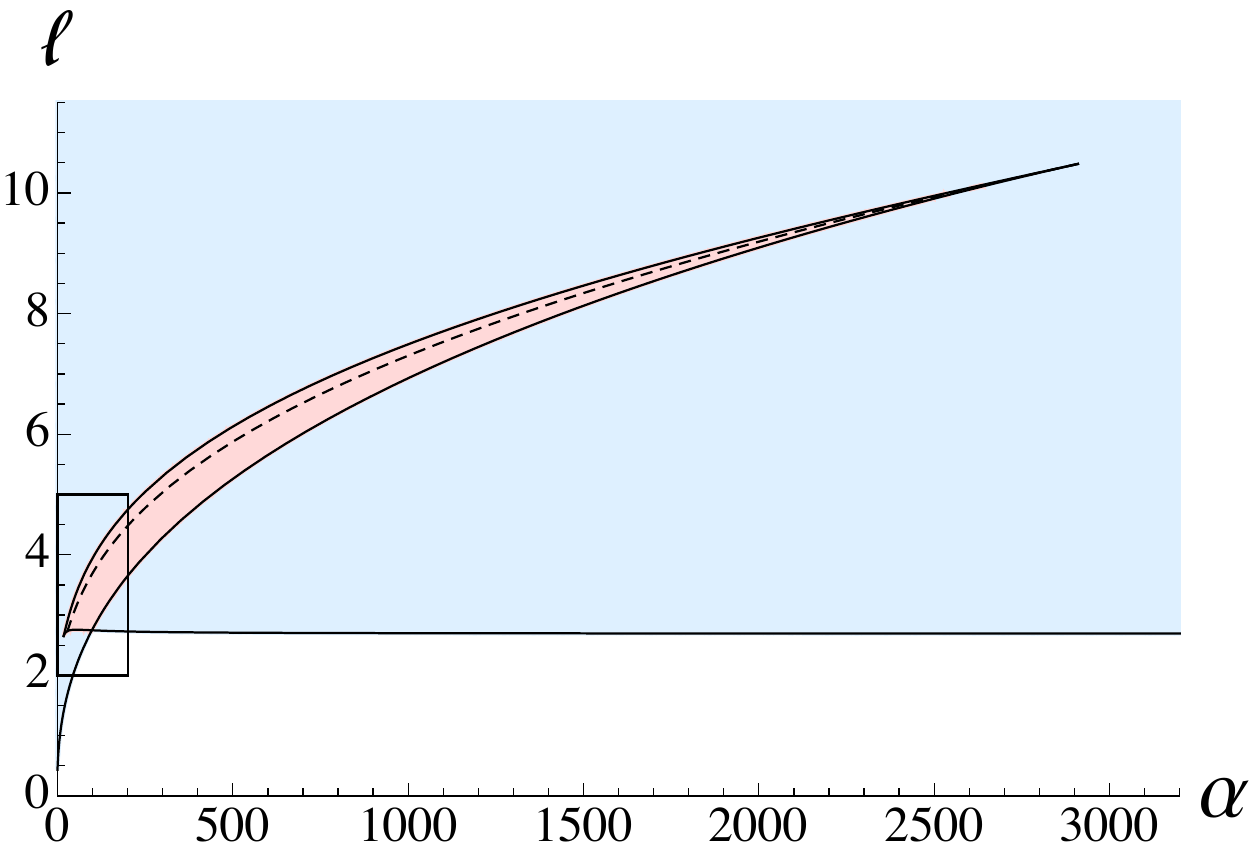}
\hspace{0.2cm}
\includegraphics[width=2.1cm]{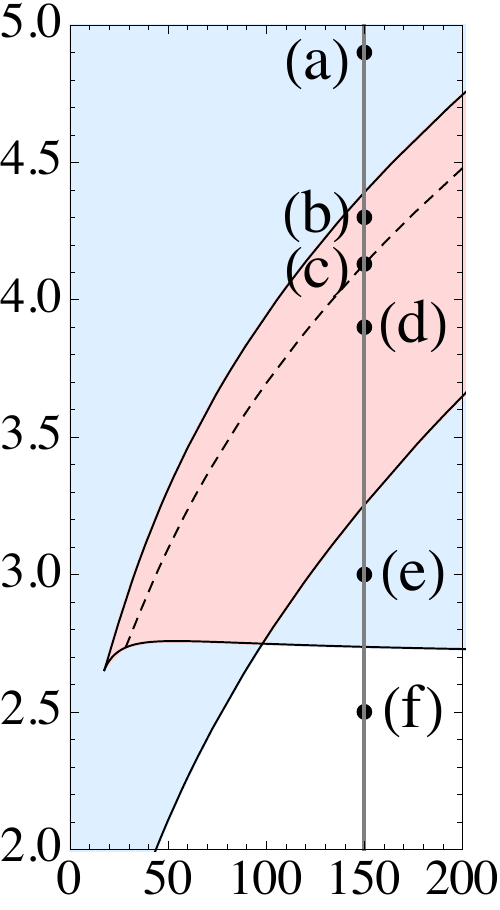}
\hspace{0.2cm}
\includegraphics[width=6cm]{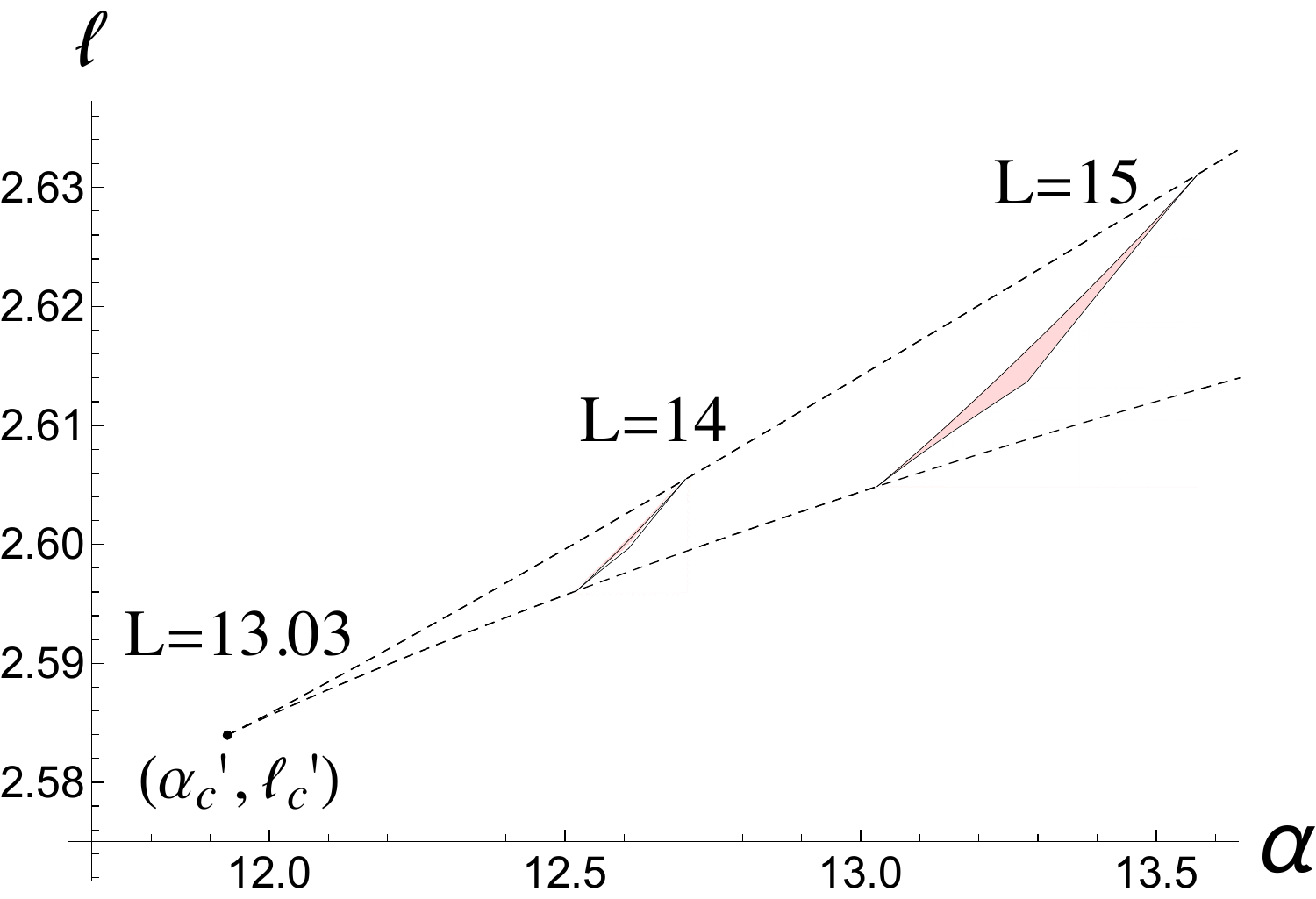}
\end{center}
\caption{ 
Left panel: number of local minima in $(\alpha,\ell)$ space for $d=6$ EGB-AdS black holes, where  we set $r_0=1$ and $L=500\, r_0$. 
Middle panel: closeup region of the box in the left panel.  
The black dots, (a)-(f)  in the middle panel correspond to the effective potentials, (a)-(f) in Fig.~\ref{fig:Veff-AdS}, respectively.
The effective potential allows only one local minimum in the blue region and two local minima in the red region, respectively. 
In the red region, the two local minima become equal values on the dashed curve [(c) in Fig.\ref{fig:Veff-AdS}], above which the outer minimum is smaller and vice versa [(b) and (d) in Fig.\ref{fig:Veff-AdS}, respectively].
The right panel shows how the red regions in $d=6$ change when $L$ changes within the range of $13.03...\le L \le  15$ for $r_0=1$, and 
the red region shrinks to a point at  $L=L_c=:13.03...$.
 \label{fig:SBOAdS}}
\end{figure}
 
We find the potential can have  either one or two local minima, which depends on the parameters $(\alpha,\ell,L)$. 
Figure~\ref{fig:SBOAdS} displays how the number of  local minima depends on the parameters $(\alpha,\ell)$ for fixed $(L,r_0)$ in the $d=6$ AdS-EGB black hole, where we set $L=500\, r_0$ as a large enough $L$. 
The two local minima are allowed in the red region, whereas only one local minimum is  allowed in the blue region. 
As $L$ becomes smaller, the red region shrinks, and eventually disappears below a certain critical value $L=L_c$, which can be obtained by solving 
\begin{align}
 V_{\rm eff}'(r_c')_{\alpha_c',\ell_c',L_c}= V_{\rm eff}''(r_c')_{\alpha_c',\ell_c',L_c}= V_{\rm eff}^{(3)}(r_c')_{\alpha_c',\ell_c',L_c}= V_{\rm eff}^{(4)}(r_c')_{\alpha_c',\ell_c',L_c}=0,
\end{align}
where it should be noted that $(\alpha_c',\ell'_c,r'_c)$ are different from $(\alpha_c,\ell_c,r_c)$ in Eq.~(\ref{eq:cusp-cond}). 
Note that $\alpha_c'$ also gives the lower bound of $\alpha$ for the allowance of two SCOs in each dimension.
The critical values $L_c$ and $\alpha_c'$ are listed in Table~\ref{table:Lcrit}. 
\begin{table}[H]
\centering
\begin{tabular}{|c||c|c|c|c|}\hline
$d$ & 6 & 7 & 8 & 9 \\\hline
$L_c/r_0$ & 13.03 & 12.52 & 13.60 & 16.89\\\hline
$\alpha_c'/r_0^2$&11.93& 16.21& 25.23& 46.66\\ \hline
\end{tabular}\label{table:Lcrit}
\caption{Critical value $L_c$ and  the corresponding $\alpha_c'$ in each dimension.  When $L\le L_c$ or $\alpha\le \alpha_c'$, the effective  potential does not have two local minima.}
\end{table}

\begin{figure}[H]
\begin{center}
\includegraphics[width=5.4cm]{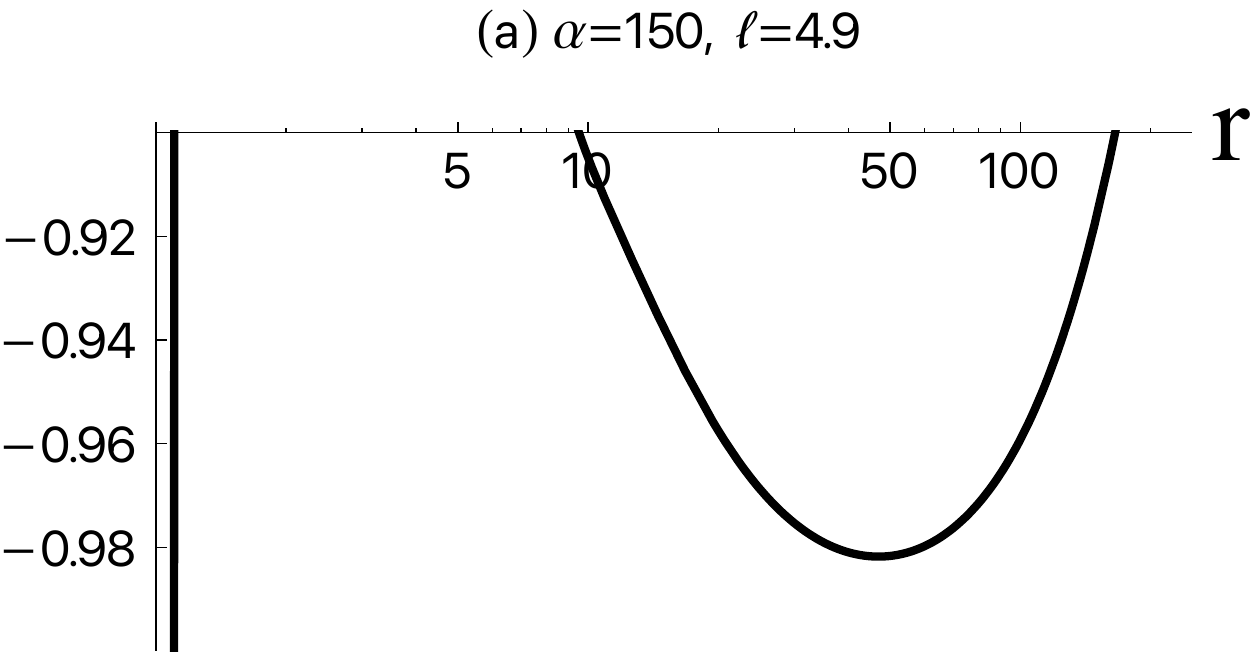}
\includegraphics[width=5.4cm]{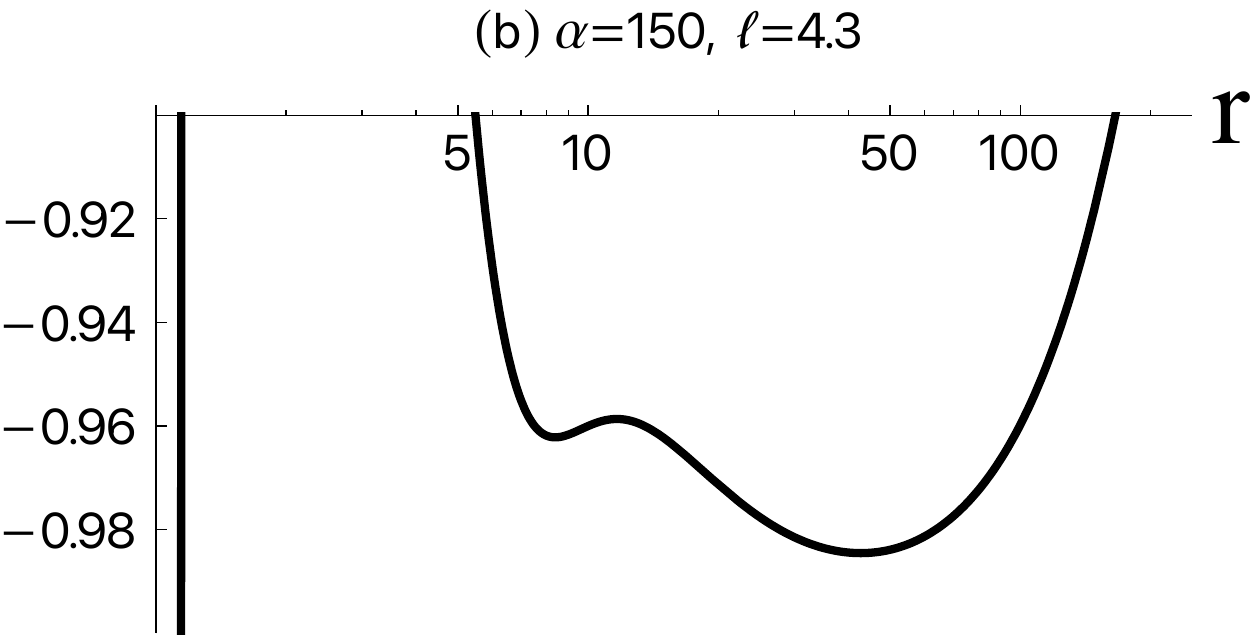}
\includegraphics[width=5.4cm]{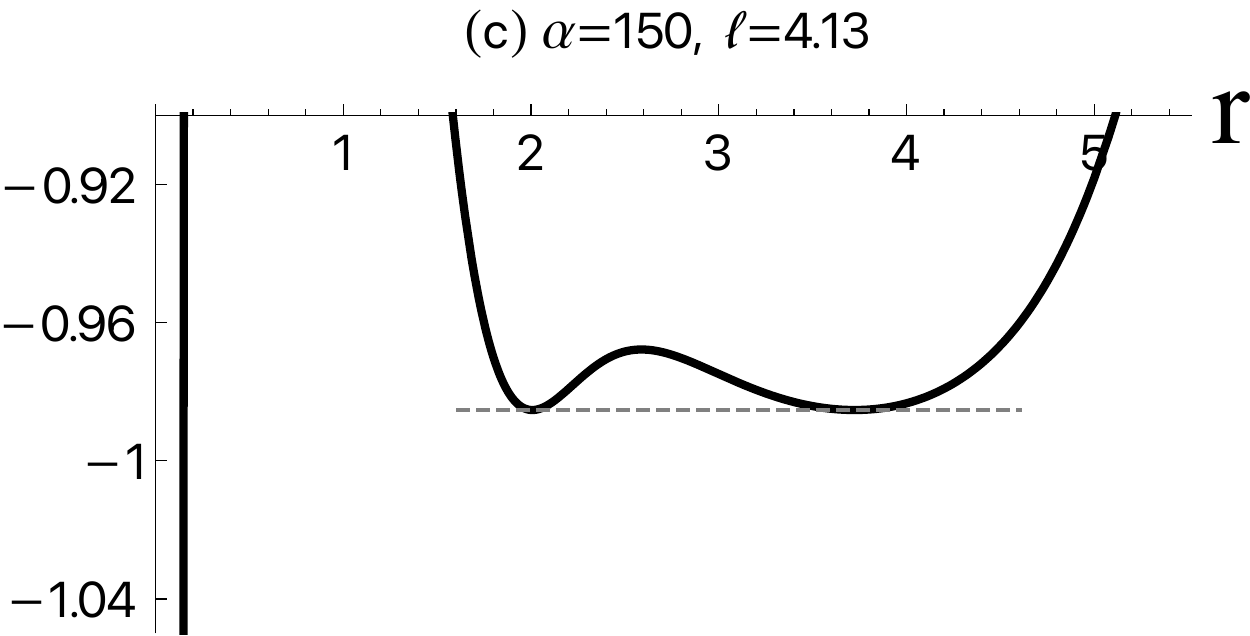}
\end{center}
\begin{center}
\includegraphics[width=5.4cm]{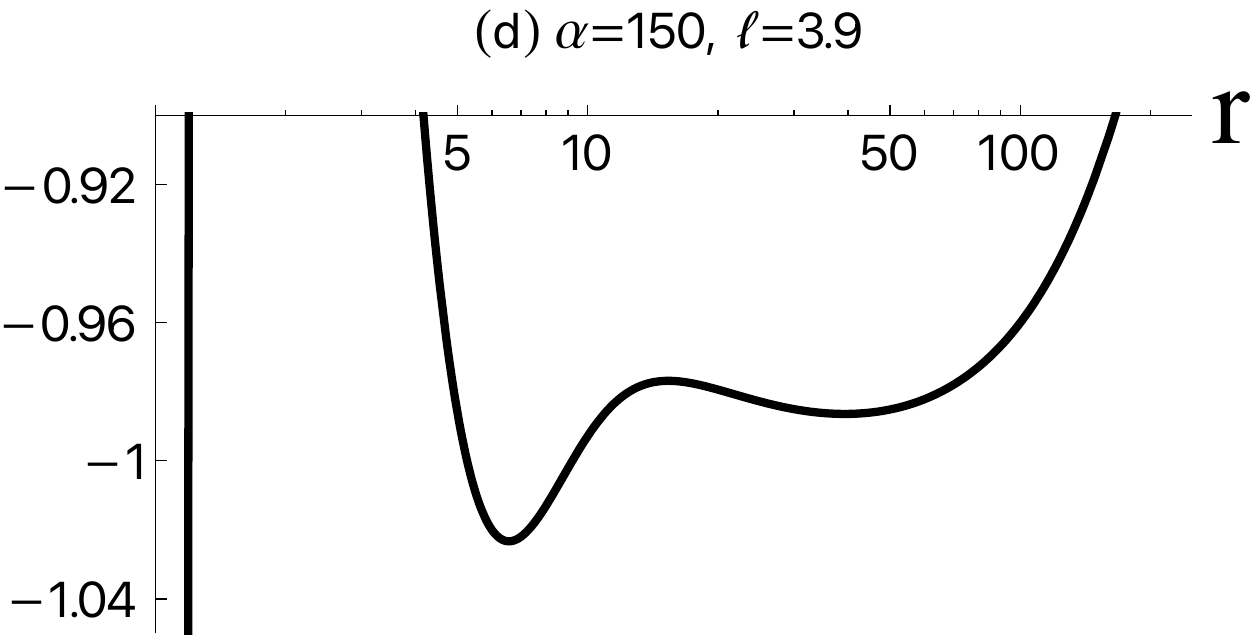}
\includegraphics[width=5.4cm]{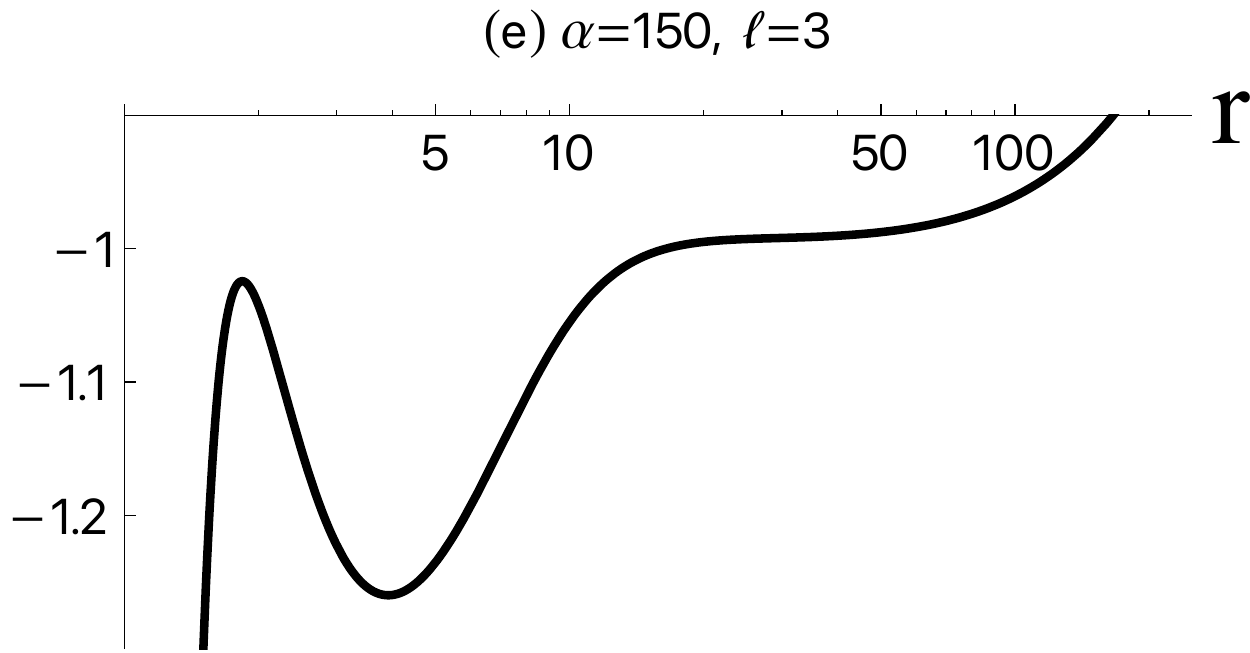}
\includegraphics[width=5.4cm]{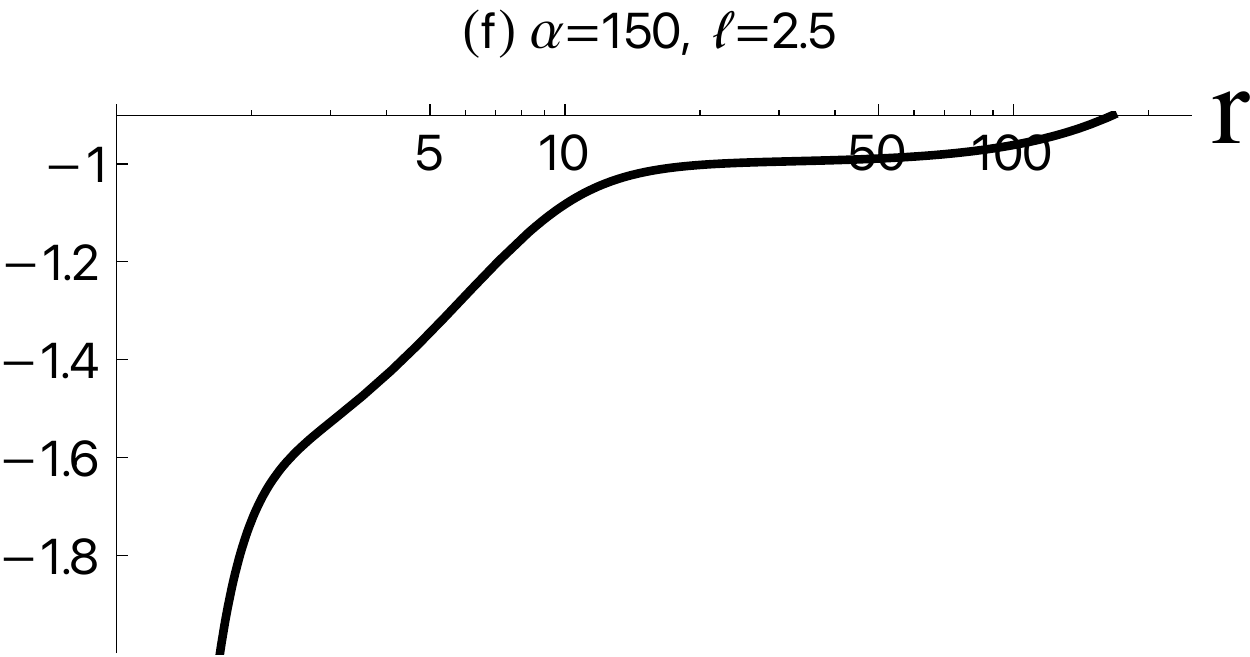}
\end{center}
\caption{Effective potentials for EGB-AdS black holes with $L=500\, r_0$ and $\alpha=150\, r_0$. In the plot, every quantity is written in units of $r_0$. The two minima have equal depth in panel (c). \label{fig:Veff-AdS}}
\end{figure}

\section{Discussion}\label{sec:sum}

In this paper, we have studied stable bound orbits around the static, spherically symmetric black holes in EGB theories with and without a negative cosmological constant. 
We have analytically shown that stable bound orbits exist only for a sufficiently large GB coupling in $6 \le d \le 9$, which was numerically   observed in~Ref.\cite{Rosa:2008dh}. 
Furthermore, we have numerically derived the parameter region of $(\alpha,\ell)$ for a fixed AdS radius in which stable  bound orbits exist outside a black hole horizon. 
As is essentially different from $6 \le d \le 8$,  the ISCO radius in $d=9$  seems to diverge at the pure GB limit $\alpha \to \infty$,  which is consistent with the nonexistence of stable bound orbits in $d=9$ pure GB theory.

\medskip
As for the AdS-EGB black holes, the interplay between the GB correction and AdS barrier can make  two local minima of the  effective potential.  
This existence of more than one stable  bound orbit is expected to admit a variety of rich dynamics.
In particular, in the context of AdS/CFT correspondence, this will also lead to some interesting physics. 
In contrast,
the dS-EGB black holes, which we have not dealt with in this paper, will qualitatively have the same properties as the vacuum EGB black holes, since the positive cosmological constant cannot make a further potential well.

\medskip
Recently, using a so-called large $D$ approach for the EGB theories~\cite{Suzuki:2022apk}, we  found the first analytic solutions of equally rotating EGB black holes in odd dimensions. 
It may be physically interesting to investigate stable  bound orbits around such rotating black holes, since a particle rotating in the same  direction as a black hole rotation behaves entirely differently  from one rotating in the opposite direction. 
Therefore, particle motion becomes more complicated around a rotating black hole. 
This analysis is the focus of our future work.       

\medskip
Our result also provides an interesting implication on the formation of higher dimensional black holes that may occur in a future collider such as Future Circular Collider, although it has not yet been confirmed at the Large Hadron Collider. 
If a EGB black hole is formed and radiates massive particles by Hawking radiation, 
these particles can be stably trapped in the potential well outside the horizon. 
This may significantly change the energy spectrum observed by a detector, compared with  the Schwarzschild-Tangherlini black holes.

\begin{acknowledgements}
We thank Takahisa Igata for useful comments.
RS was supported by JSPS KAKENHI Grant Number~JP18K13541.
ST was supported by JSPS KAKENHI Grant Number~17K05452 and 21K03560.
\end{acknowledgements}


\begin{thebibliography}{99}
\bibitem{Wilkins:1972rs}
D.~C.~Wilkins,
``Bound Geodesics in the Kerr Metric,''
Phys. Rev. D \textbf{5}, 814-822 (1972)





\bibitem{Cardoso:2014sna}
V.~Cardoso, L.~C.~B.~Crispino, C.~F.~B.~Macedo, H.~Okawa and P.~Pani,
Light rings as observational evidence for event horizons: Long-lived modes, ergoregions and nonlinear instabilities of ultracompact objects,
Phys. Rev. D \textbf{90}, 
044069 (2014)
[arXiv:1406.5510 [gr-qc]].





\bibitem{Khoo:2016xqv}
F.~S.~Khoo and Y.~C.~Ong,
``Lux in obscuro: Photon Orbits of Extremal Black Holes Revisited,''
Class. Quant. Grav. \textbf{33}, no.23, 235002 (2016)
[erratum: Class. Quant. Grav. \textbf{34}, no.21, 219501 (2017)]
[arXiv:1605.05774 [gr-qc]].




\bibitem{Dolan:2016bxj}
S.~R.~Dolan and J.~O.~Shipley,
Stable photon orbits in stationary axisymmetric electrovacuum spacetimes,
Phys. Rev. D \textbf{94}, 
044038 (2016)
[arXiv:1605.07193 [gr-qc]].


\bibitem{Nakashi:2019mvs}
K.~Nakashi and T.~Igata,
Innermost stable circular orbits in the Majumdar-Papapetrou dihole spacetime,
Phys. Rev. D \textbf{99}, 
124033 (2019)
[arXiv:1903.10121 [gr-qc]].

\bibitem{Nakashi:2019tbz}
K.~Nakashi and T.~Igata,
Effect of a second compact object on stable circular orbits,
Phys. Rev. D \textbf{100}, 
104006 (2019)
[arXiv:1908.10075 [gr-qc]].


\bibitem{Keir:2014oka}
J.~Keir,
Slowly decaying waves on spherically symmetric spacetimes and ultracompact neutron stars,
Classical Quantum Gravity \textbf{33}, 
135009 (2016)
[arXiv:1404.7036 [gr-qc]].



\bibitem{Tangherlini:1963bw} 
  F.~R.~Tangherlini,
  Schwarzschild field in $n$ dimensions and the dimensionality of space problem,
  Nuovo Cimento \textbf{27}, 636 (1963).


\bibitem{Page:2006ka} 
  D.~N.~Page, D.~Kubiznak, M.~Vasudevan and P.~Krtous,
  Complete Integrability of Geodesic Motion in General Kerr-NUT-AdS Spacetimes,
  Phys.\ Rev.\ Lett.\ \textbf{98}, 061102 (2007)
  [arXiv:hep-th/0611083].
    
\bibitem{Frolov:2003en} 
  V.~P.~Frolov and D.~Stojkovic,
  Particle and light motion in a space-time of a five-dimensional rotating black hole,
  Phys.\ Rev.\ D \textbf{68}, 064011 (2003)
  [arXiv:gr-qc/0301016].
    
\bibitem{Frolov:2006pe} 
  V.~P.~Frolov, P.~Krtous and D.~Kubiznak,
  Separability of Hamilton-Jacobi and Klein-Gordon equations in general Kerr-NUT-AdS spacetimes,
  J.\ High Energy Phys. 02 (2007) 005.
  [arXiv:hep-th/0611245].
  
  
  \bibitem{Cardoso:2008bp}
  V.~Cardoso, A.~S.~Miranda, E.~Berti, H.~Witek and V.~T.~Zanchin,
  Geodesic stability, Lyapunov exponents and quasinormal modes,
  Phys.\ Rev.\ D \textbf{79}, 064016 (2009) 
  [arXiv:0812.1806 [hep-th]].
  

 


\bibitem{Igata:2010ye} 
  T.~Igata, H.~Ishihara and Y.~Takamori,
  Stable bound orbits around black rings,
  Phys.\ Rev.\ D \textbf{82}, 101501(R) (2010)
  [arXiv:1006.3129 [hep-th]].
     
 
\bibitem{Igata:2010cd}
  T.~Igata, H.~Ishihara and Y.~Takamori,
  Chaos in geodesic motion around a black ring,
  Phys.\ Rev.\ D \textbf{83}, 047501 (2011)
  [arXiv:1012.5725 [hep-th]].
  
  
\bibitem{Igata:2013be} 
  T.~Igata, H.~Ishihara and Y.~Takamori,
  Stable bound orbits of massless particles around a black ring,
  Phys.\ Rev.\ D \textbf{87}, 104005 (2013)
  [arXiv:1302.0291 [hep-th]].
 










\bibitem{Tomizawa:2019egx}
S.~Tomizawa and T.~Igata, 
Stable bound orbits around a supersymmetric black lens,
Phys.\ Rev.\ D \textbf{100}, 124031 (2019)
[arXiv:1908.09749 [hep-th]].


\bibitem{Tomizawa:2020mvw}
S.~Tomizawa and T.~Igata,
``Stable bound orbits in black lens backgrounds,''
Phys. Rev. D \textbf{102}, 124079 (2020)
[arXiv:2011.11002 [hep-th]].




\bibitem{Igata:2020vlx}
T.~Igata and S.~Tomizawa,
Stable circular orbits in higher-dimensional multi-black hole spacetimes,
Phys. Rev. D \textbf{102},  084003 (2020)
[arXiv:2008.00179 [hep-th]].




\bibitem{Igata:2021wwj}
T.~Igata and S.~Tomizawa,
``Stable circular orbits in caged black hole spacetimes,''
Phys. Rev. D \textbf{103}, no.8, 084011 (2021)
[arXiv:2102.00800 [gr-qc]].


\bibitem{Tomizawa:2021vaa}
S.~Tomizawa and T.~Igata,
``Stable circular orbits in Kaluza-Klein black hole spacetimes,''
Phys. Rev. D \textbf{103}, no.12, 124004 (2021)
[arXiv:2103.08581 [hep-th]].






\bibitem{Tomizawa:2022kpt}
S.~Tomizawa and R.~Suzuki,
``Stable bound orbits in microstate geometries,''
[arXiv:2203.15185 [hep-th]].



\bibitem{Eperon:2016cdd} 
  F.~C.~Eperon, H.~S.~Reall and J.~E.~Santos,
  Instability of supersymmetric microstate geometries,
  J.\ High Energy Phys.\ 10 (2016) 031
  [arXiv:1607.06828 [hep-th]].
  
 
\bibitem{Eperon:2017bwq}
F.~C.~Eperon,
``Geodesics in supersymmetric microstate geometries,''
Class. Quant. Grav. \textbf{34}, no.16, 165003 (2017)
[arXiv:1702.03975 [gr-qc]].





\bibitem{Dadhich:2013moa}
N.~Dadhich, S.~G.~Ghosh and S.~Jhingan,
``Bound orbits and gravitational theory,''
Phys. Rev. D \textbf{88}, no.12, 124040 (2013)
doi:10.1103/PhysRevD.88.124040
[arXiv:1308.4770 [gr-qc]].



\bibitem{Dadhich:2021vdd}
N.~Dadhich and S.~Shaymatov,
``Circular orbits around higher dimensional Einstein and pure Gauss-Bonnet rotating black holes,''
[arXiv:2104.00427 [gr-qc]].


\bibitem{Boulware:1985wk}
D.~G.~Boulware and S.~Deser,
``String Generated Gravity Models,''
Phys. Rev. Lett. \textbf{55}, 2656 (1985).



\bibitem{Bhawal:1990nh}
B.~Bhawal,
``Geodesics in Boulware-Deser black hole space-time,''
Phys. Rev. D \textbf{42}, 449-452 (1990)



\bibitem{Rosa:2008dh}
V.~M.~Rosa and P.~S.~Letelier,
``Circular Orbits in Einstein-Gauss-Bonnet Gravity,''
Phys. Rev. D \textbf{78}, 084038 (2008)
[arXiv:0810.1177 [gr-qc]].


\bibitem{Maeda:2006pm}
H.~Maeda,
``Final fate of spherically symmetric gravitational collapse of a dust cloud in Einstein-Gauss-Bonnet gravity,''
Phys. Rev. D \textbf{73}, 104004 (2006)
[arXiv:gr-qc/0602109 [gr-qc]].



\bibitem{Konoplya:2020ptx}
R.~A.~Konoplya and A.~Zhidenko,
``Massive particles in the Einstein-Lovelock-anti-de Sitter black hole spacetime,''
Class. Quant. Grav. \textbf{38}, no.4, 045015 (2021)
[arXiv:2010.09064 [gr-qc]].

\bibitem{Suzuki:2022apk}
R.~Suzuki and S.~Tomizawa,
``Rotating black holes at large $D$ in Einstein-Gauss-Bonnet theory,''
[arXiv:2202.12649 [hep-th]].










\end{thebibliography}
\end{document}